\RequirePackage{fix-cm}

\documentclass[twocolumn]{svjour3}          % twocolumn
\usepackage{lipsum}
\smartqed   %flush right qed marks, e.g. at end of proof
\usepackage{graphicx}
\usepackage{amsmath}
\usepackage{amsfonts}
\begin{document} \sloppy
%\lipsum[1-14]
\title{A coarse grid projection method for accelerating heat transfer computations 
%Transport of passive scalars in incompressible flows over a cylinder using a coarse-grid projection method
}



\author{A. Kashefi     %    \and
       % A. E. Staples %etc.
}


\institute{A. Kashefi \at
              Department of Mechanical Engineering, Stanford University, Stanford, CA 94305, USA \\
  %            Tel.: +123-45-678910\\
    %          Fax: +123-45-678910\\
              \email{kashefi@stanford.edu}             \\
  %         \and
           %A. E. Staples \at
           %   Faculty of Engineering, Engineering Science and Mechanics Program, Virginia Tech, Blacksburg, VA 24061, USA \\
  %            Tel.: +123-45-678910\\
    %          Fax: +123-45-678910\\
          %    \email{aestaples@vt.edu}           %  \\
}

\date{Received: date / Accepted: date}

\maketitle

\begin{abstract}
Coarse Grid Projection (CGP) methodology is used to accelerate the computations of sets of decoupled nonlinear evolutionary and linear static equations. In CGP, the linear equations are solved on a coarsened mesh compared to the nonlinear equations, leading to a reduction in central processing unit (CPU) time. The accuracy of the CGP scheme has been assessed for the advection-diffusion equation along with the pressure Poisson equation. Here we add another decoupled equation to this set: the energy equation. In this article, we examine the influence of CGP methodology for the first time on thermal fields. To this purpose, a semi-implicit-time-integration unstructured-triangular-finite-element CGP version is selected. The CGP platform is validated with two different test cases: first, natural convection induced by a hot circular cylinder located in the center of a cold square cylinder, and second, the flow over a circular cylinder with the condition of constant cylinder temperature. Regarding the first test case, the CGP and non-CGP simulations are carried out for different Rayleigh numbers. The velocity and temperature fields as well as the local Nusselt number on the surface of the inner hot cylinder calculated by CGP reveal good agreement with the non-CGP data. Concerning the second test case, the temperature variable is used as the passive scalar. For different Prandtl numbers, we compare the CGP and non-CGP configurations according to the Nusselt number and the spatial structure of the scalar field obtained. The phase lag between the standard and CGP approaches is transmitted from the velocity field into the temperature filed, and thus into the local transient Nusselt number. For one and two levels of coarsening, the numerical predictions by CGP for the unsteady local heat transfer coefficients agree well with available data in the literature. In general, CGP is able to maintain excellent to reasonable accuracy of the temperature filed, while achieves speedup factors ranged approximately from 1.7 to 3.7. 
\
\
\

%\PACS{PACS code1 \and PACS code2 \and more}
\
\
\
\
\
\

%\textbf{Mathematics Subject Classification (2010)} {35Q30 \and 65Y20 \and 65N30 %\and 65N55}
%\subclass{35Q30 \and 65Y20\and 65N30 \and 65N55}
\end{abstract}

\section{Introduction}
\label{intro}

Pressure projection schemes are widely used for the unsteady incompressible flow computations [1--4]. Taking the advantages of these techniques, the saddle-point issue of the continuity and momentum equations disappears [3, 4]. Hence, one only deals with two decoupled cascading elliptic equations: the advection-diffusion equation and the pressure Poisson one. Different multigrid schemes have been already introduced to lessen the computational times associated with the numerical pressure correction methods (see e.g., Refs. [5--9]). Coarse Grid Projection (CGP) methodology is a recently used multiresolution scheme to accelerate these computations [10--13]. CGP saves a considerable amount of CPU time by reducing the degree of freedom for the discretized Poisson equation, which is the most time consuming subproblem. Accordingly, the nonlinear advection-diffusion equation is solved on a fine grid and the linear pressure Poisson equation is solved on a corresponding coarsened grid. Mapping functions transfer data between the grids. The CGP procedure is described in detail in Sect. 2.2.

In 2010 Lentine et al. [10] first introduced CGP for accelerating inviscid flow computations. In 2013 San and Staples [11] used the CGP technique for the numerical simulation of the incompressible Navier-Stokes equations. In 2014 the CGP algorithm was used in the fast fluid dynamics (FFD) models by Jin and Chen [12]. In 2018 a finite element version of CGP with a semi-implicit time integration scheme was presented by Kashefi and Staples [13]. In 2019 Kashefi [14] discussed CGP as a guide for partial mesh refinement of incompressible flow computations.

In all the literature cited above, the authors [10--14] studied the performance of CGP in terms of the level of accuracy obtained in the velocity or pressure fields and achieved speedup factors. Nonetheless, the influence of the CGP algorithm on the energy equation has not yet been investigated.

The study of the energy equation in a numerical simulation performed by the CGP technique is important in two aspects. First, since the advection term in the energy equation is based on the velocity field obtained by CGP, preserving the accuracy level of the thermal field should be investigated. Second, in order to obtain the velocity and thermal fields using pressure projection schemes, one has to deal with three decoupled elliptic equations at each time step: a linearized equation for the intermediate velocity field, a linear Poisson equation for the pressure field, and a linearized equation for the thermal field. Hence, the contribution of the CGP scheme to accelerating the computations becomes significant. To this end, we consider two different practical situations.

First, when the buoyancy force leads to the thermally-driven flows with the so-called Boussinesq approximation. By this assumption, the solution of the energy equation appears in the momentum balance as the source term [15]. The natural convection in a square enclosure with a circular cylinder for different Rayleigh numbers is considered as a standard test case for this condition.

Second, for small scalar differences, the energy equation is, indeed, the conservation equation of a passive scalar and can be independently solved for a given velocity field [15]. An external unsteady flow past a cylinder is a physically meaningful benchmark case and a model for canonical studies of demanding fluid mechanics problems [17]. Thus, this test case is solved for different Prandtl numbers in order to investigate the performance of the CGP strategy for the study of transporting passive scalars.

The rest of this article is structured as follows. Section 2.1 gives the governing equations for incompressible flows and conservation of energy. We discuss coarse-grid projection methodology in Sect. 2.2. Computational aspects of the problem are described in Sect. 2.3. Numerical results and their relevant discussions are presented in Sect. 3. Conclusions and notes for extensions of the work are provided in Sect. 4. 

\section{Problem formulation}
\subsection{Governing equations}
\label{sec:2}
The equations of conservation of momentum, mass, and energy for an incompressible flow of a Newtonian fluid are given by
\begin{equation}
\rho \bigg[\frac{\partial \textbf{\textit{u}}}{\partial t}+(\textbf{\textit{u}}\cdot \nabla)\textbf{\textit{u}}\bigg] -\mu \Delta \textbf{\textit{u}} + \nabla p=\textbf{\textit{f}} \textrm{ in } V,
\end{equation}
\begin{equation}
\nabla \cdot \textbf{\textit{u}}=0 \textrm{ in } V,
\end{equation}
\begin{equation}
\textbf{\textit{u}}=\textbf{\textit{u}}_{\Gamma_D} \textrm{ on } \Gamma_D,
\end{equation}
\begin{equation}
-p\textbf{\textit{n}}+\mu \nabla \textbf{\textit{u}} \cdot \textbf{\textit{n}}=\textbf{\textit{t}}_{\Gamma_N} \textrm{ on } \Gamma_N,
\end{equation}
\begin{equation}
\rho \bigg[\frac{\partial\theta}{\partial t}+(\textbf{\textit{u}}\cdot \nabla)\theta\bigg]=\frac{\kappa}{c_p}\Delta\theta \textrm{ in } V,
\end{equation}
\begin{equation}
\theta=\theta_{\Omega_D} \textrm{ on } \Omega_D,
\end{equation}
\begin{equation}
\nabla \theta \cdot \textbf{\textit{n}}=\textbf{\textit{b}}_{\Omega_N} \textrm{ on } \Omega_N,
\end{equation}
where $\textbf{\textit{u}}$  is the velocity vector, $p$ stands for the pressure, and $\theta$ represents the temperature of the fluid in domain $V$. $\textbf{\textit{f}}$ is the vector of external force. $\textbf{\textit{t}}_{\Gamma_N}$ and $\textbf{\textit{b}}_{\Omega_N}$ denote the stress vectors applied to the velocity and temperature fields, respectively. $\rho$ is the fluid density and $\mu$ is the dynamic viscosity. $\kappa$ is the conductivity of the fluid and $c_p$ is the specific heat at a constant pressure. $\Gamma_D$ and $\Gamma_N$ respectively represent the velocity Dirichlet and Neumann boundaries, while   $\Omega_D$ and $\Omega_N$ respectively denote the temperature Dirichlet and Neumann boundaries of the domain $V$. $\textbf{\textit{n}}$ is the outward unit vector normal to the boundaries. There is no overlapping between $\Gamma_D$ and $\Gamma_N$ subdomains. Similarly no overlap exists between $\Omega_D$ and $\Omega_N$ subdomains as well.

We discretize the system of equations using a first-order semi-implicit time integration formula [18]. Then, we apply a non-incremental pressure correction scheme [4] to the time-discretized system, yielding to

\begin{equation}
\rho \bigg[\frac{\textbf{\textit{\~u}}^{n+1} - \textbf{\textit{u}}^n}{\delta t}+(\textbf{\textit{u}}^n\cdot \nabla)\textbf{\textit{\~u}}^{n+1}\bigg] -\mu \Delta \textbf{\textit{\~u}}^{n+1}=\textbf{\textit{f}}^{n} \textrm{ in } V,
\end{equation}

\begin{equation}
\textbf{\textit{\~u}}^{n+1}=\textbf{\textit{u}}_{\Gamma_D}^{n+1} \textrm{ on } \Gamma_D,
\end{equation}

\begin{equation}
\mu \nabla \textbf{\textit{\~u}}^{n+1} \cdot \textbf{\textit{n}}=\textbf{\textit{t}}_{\Gamma_N}^{n+1} \textrm{ on } \Gamma_N,
\end{equation}

\begin{equation}
\Delta p^{n+1}=\frac{\rho}{\delta t} \nabla \cdot \textbf{\textit{\~u}}^{n+1} \textrm{ in } V,
\end{equation}

\begin{equation}
\nabla p^{n+1} \cdot\textit{\textbf{n}}=0  \textrm{ on } \Gamma_D,
\end{equation}

\begin{equation}
p^{n+1}=0  \textrm{ on } \Gamma_N,
\end{equation}

\begin{equation}
\textbf{\textit{u}}^{n+1}=\textbf{\textit{\~u}}^{n+1}-\frac{\delta t}{\rho}\nabla p^{n+1}\textrm{ in } V,
\end{equation}

\begin{equation}
\rho \bigg[\frac{\theta^{n+1} - \theta^n}{\delta t}+(\textbf{\textit{u}}^{n+1}\cdot \nabla)\theta^{n+1}\bigg] =\frac{\kappa}{c_p}\Delta\theta^{n+1} \textrm{ in } V,
\end{equation}

\begin{equation}
\theta^{n+1}=\theta_{\Omega_D}^{n+1} \textrm{ on } \Omega_D,
\end{equation}

\begin{equation}
\nabla \theta^{n+1} \cdot \textbf{\textit{n}}=\textbf{\textit{b}}_{\Omega_N}^{n+1} \textrm{ on } \Omega_N,
\end{equation}
where $\delta t$ represents the time step and $\textbf{\textit{\~u}}$ is the intermediate velocity vector. For a more detailed description of the pressure projection scheme implemented here, one may refer to Refs. [1--4].

The finite-element Galerkin scheme [3, 19] with the piecewise linear basis function $\textbf{P}_1$ is used to spatially discretize the space of the velocity, pressure, and temperature fields. The finite-element form of Eqs. (8)--(17) is expressed as

\begin{equation}
\frac{1}{\delta t}\big(\textbf{M}_v\textrm{\~U}^{n+1}
-\textbf{M}_v\textrm{U}^{n}\big)+\big[\textbf{N}^{n}+\textbf{L}_v\big]\textrm{\~U}^{n+1}=\textbf{M}_v\textrm{F}^{n},
\end{equation}

\begin{equation}
\textbf{L}_p\textrm{P}^{n+1}=\frac{\rho}{\delta t}\textbf{D}\textrm{\~U}^{n+1},
\end{equation}

\begin{equation}
\textbf{M}_v\textrm{U}^{n+1}=\textbf{M}_v\textrm{\~U}^{n+1}-\delta
t\textbf{G}\textrm{P}^{n+1},
\end{equation}

\begin{equation}
\frac{1}{\delta t}\big(\textbf{M}_\theta\Theta^{n+1}
-\textbf{M}_\theta\Theta^{n}\big)+\big[\textbf{N}^{n+1}+\textbf{L}_\theta\big]\Theta^{n+1}=\textrm{Q}^{n+1},
\end{equation}
where \textbf{M}$_v$, \textbf{M}$_\theta$, \textbf{L}$_v$, \textbf{L}$_p$, \textbf{L}$_\theta$, \textbf{D}, and \textbf{G} denote the matrices associated, respectively, to the velocity mass, temperature mass, velocity laplacian, pressure laplacian, temperature laplacian, divergence, and gradient operators. \textbf{N}$^{n}$ and \textbf{N}$^{n+1}$ indicate the advection operators at time $t^{n}$ and $t^{n+1}$, respectively. The vectors  {\~U}, {U}, ${\Theta}$,  {P}, {F}, and {Q} represent the nodal values of the intermediate velocity, the end-of-step velocity, the temperature, the pressure, the forcing term on the velocity domain, and the stress term on the temperature domain, respectively.

\subsection{Coarse grid projection methodology}
\label{sec:3}

In the CGP scheme, first we balance the advection-diffusion equation on a fine grid and obtain the intermediate velocity field data \~{U}$_f^{n+1}$. Then, we restrict \~{U}$_f^{n+1}$ to a corresponding coarsened grid and set  \~{U}$_c^{n+1}$. We take the divergence of the restricted intermediate velocity \~{U}$_c^{n+1}$ in order to set the source term of the pressure Poisson equation. We solve the Poisson equation on the coarsened grid and obtain P$_c^{n+1}$. In the next stage, we prolong the resulting pressure data P$_c^{n+1}$ from the coarse grid to the fine grid and set P$_f^{n+1}$. We correct the velocity domain and obtain U$_f^{n+1}$ on the fine grid. Now, we create the advection matrix \textbf{N}$^{n+1}$ based to the obtained velocity field data U$_f^{n+1}$. Finally, we solve the last conservation equation for the energy on the fine grid to obtain ${\Theta}_f^{n+1}$.

In practice we consider four nested spaces: $V_1 \subset V_2 \subset V_3 \subset V_4=V$. We uniformly subdivide each triangular element of the discretized space of $V_l$ (for 1$\leq l \leq$3) into four triangles. This procedure provides the discretized space of $V_{l+1}$. Hence, for each CGP simulation we have a fine mesh and a corresponding coarsened mesh respectively with $M$ and $N$ elements such that $N=4^{-k}M$, where $k$ indicates the coarsening level. The restriction $R:V_4 \rightarrow V_{4-l}$ and prolongation $P:V_l \rightarrow V_{l+1}$ operators and their matrix representations, respectively $\textbf{R}_{\textbf{4}}^{\textbf{4-l}}$ and $\textbf{P}_{\textbf{l}}^{\textbf{l+1}}$, are constructed using Geometric Multigrid (GMG) tools. $\textbf{R}_{\textbf{4}}^{\textbf{4-l}}$ injects the intermediate velocity data from a fine grid ($V_4$) into the corresponding coarse grid ($V_{4-l}$). $\textbf{P}_{\textbf{l}}^{\textbf{l+1}}$ corresponds to the finite-element shape functions. Since we implement $\textbf{P}_1$ in this study, $\textbf{P}_{\textbf{l}}^{\textbf{l+1}}$ prolongs the pressure data from the coarse grid ($V_l$) to the next nested space ($V_{l+1}$) using a linear interpolation. Finally, we derive the pressure laplacian $\bar{\textbf{L}}_p$ and divergence $\bar{\textbf{D}}$ operators on a relatively coarse mesh ($V_{4-l}$) by taking the inner products of the coarse grid finite-element shape functions. One may see Sect. 2.3 of Ref. [13] for further details.

Eqs. (22)--(27) summarize the CGP algorithm at each time step, $\delta t$, of the simulation.   \\
1.	Calculate {\~U}$_f^{n+1}$ on $V$ by solving
\begin{equation}
\big(\textbf{M}_v+\delta t \textbf{N}^{n} + \delta t \textbf{L}_v\big)\textrm{\~U}_f^{n+1}=\delta t\textbf{M}_v\textrm{F}^{n}+\textbf{M}_v\textrm{U}_f^{n}.
\end{equation}
2.	Map {\~U}$_f^{n+1}$ onto $V_{4-l}$ and obtain {\~U}$_c^{n+1}$ via
\begin{equation}
\textrm{\~U}_c^{n+1}=\textbf{R}_\textbf{4}^{\textbf{4-l}}\textrm{\~U}_f^{n+1}.
\end{equation}
3.	Calculate P$_c^{n+1}$ on $V_{4-l}$ by solving
\begin{equation}
\bar{\textbf{L}}_p\textrm{P}_c^{n+1}=\frac{\rho}{\delta t}\bar{\textbf{D}}\textrm{\~U}_c^{n+1}.
\end{equation}
4.	Remap P$_c^{n+1}$ onto $V$ and obtain P$_f^{n+1}$ via
\begin{equation}
\textrm{P}_f^{n+1}=\textbf{P}_\textbf{l}^{\textbf{l+1}}\textrm{P}_c^{n+1}.
\end{equation}
5.	Calculate U$_f^{n+1}$ via
\begin{equation}
\textbf{M}_v\textrm{U}_f^{n+1}=\textbf{M}_v\textrm{\~U}_f^{n+1}-\delta
t\textbf{G}\textrm{P}_f^{n+1}.
\end{equation}
6. Build up the advection operator \textbf{N}$^{n+1}$ using the obtained velocity field U$_f^{n+1}$.
\\
7.	Calculate $\Theta_f^{n+1}$ on $V$ by solving
\begin{equation}
\big(\textbf{M}_\theta+\delta t \textbf{N}^{n+1} + \delta t \textbf{L}_\theta\big)\Theta_f^{n+1}= \delta t\textrm{Q}^{n+1}+\textbf{M}_\theta\Theta_f^{n}.
 \end{equation}

\subsection{Computational consideration}
\label{sec:4}
We employ an in-house \verb C++  object oriented code. We use the $ILU (0)$ preconditioned GMRES($m$) algorithm [20, 21] to solve Eqs. (22), (24), and (27). The Gmsh application [22] is used for generating unstructured finite element meshes. All simulations are performed on a single Intel(R) Xeon(R) processor with 2.66 GHz clock rate and 64 Gigabytes of RAM.

%\subsection{Subsection title}
%\label{sec:2}
%as required. Don't forget to give each section
%and subsection a unique label (see Sect.~\ref{sec:1}).

%Text with citations \cite{Ref3} and \cite{RefJ}.

\section{Results and discussion}
\label{sec:1}

To assess the performance of the CGP configuration, two standard test cases are investigated: The natural convection in a square enclosure with a circular cylinder and transport of passive scalars in flows over a circular cylinder. Note that each subproblem has own Nusselt number definition. Notations in the form of $M:N$ demonstrate the grid resolutions of the advection-diffusion and energy equation solvers, $M$ elements, and the pressure Poisson equation solver, $N$ elements. 

\subsection{Natural convection in a square enclosure with a circular cylinder}
\label{sec:2}

One goal of the thermally-driven flow problem is an investigation of the effects of the temperature solution on the velocity field in the CGP framework. A second goal of this test case is to check the capability of the CGP method for heat transfer in complex geometries. To perform this test case, the geometry and boundary conditions are accorded to Lee et al. [15].  In this way, an opportunity for validation of our results is provided. 

A rectangular computational field $V:=[0, L]\times[0, L]$ is considered. A circular cylinder with the diameter $b$ is located at the center of the domain such that $b=0.4 L$. No-slip conditions are imposed at all the boundaries. For the temperature, $\theta=\theta_h$ is imposed at the cylinder surface, while $\theta=\theta_c$ is enforced at the remaining surfaces. According to the Boussinesq approximation [15], the forcing term is given by
\begin{equation}
\textbf{\textit{f}}^n=\rho g\beta(\theta^n - \theta_{ref})\hat{j},
\end{equation}
where $g$ indicates the magnitude of the gravity acceleration. $\beta$ is the thermal expansion. $\theta_{ref}$ stands for the reference temperature. $\hat{j}$ is the unit vector in the direction of $y$ axis. The Rayleigh number is expressed as
\begin{equation}
Ra=\frac{\rho^2 c_p g \beta (\theta_h - \theta_c) L^3}{\kappa \mu}.
\end{equation}
The local Nusselt number on the cylinder surface is determined as
\begin{equation}
Nu_\varphi=\frac{-b}{\theta_h - \theta_c}\frac{\partial \theta}{\partial \textbf{\textit{n}}}\rvert_\varphi,
\end{equation}
where $\varphi$ is the angle from the negative $x$-axis indicating the position of a point located on the cylinder surface.

The density ($\rho$), specific heat ($c_p$), hot temperature ($\theta_h$), box length ($L$), gravity acceleration ($g$), and thermal expansion ($\beta$) are set to 1.00; and the cold temperature ($\theta_c$) and reference temperature ($\theta_{ref}$) are set to 0.00 in the International Unit System. We consider different viscosity ($\mu$) and conductivity ($\kappa$) values to set the Rayleigh number. A constant time step of $\delta t=0.01$ s is selected and the numerical computations are executed until the following criterion is satisfied
\begin{equation}
\max (\frac{||\textrm{U}^{n+1}||_2 - ||\textrm{U}^n||_2}{||\textrm{U}^n||_2},\frac{||\Theta^{n+1}||_2 - ||\Theta^n||_2}{||\Theta^n||_2}) < 10^{-7},
\end{equation}
where $||...||_2$ is the $L^2$ norm.

Figure 1 illustrates the grids utilized by the Poisson solver for the Rayleigh number of $Ra=10^5$ for the standard resolution ($k=0$), one level coarsening ($k=1$), and two levels ($k=2$) coarsening. It should be noted that we use grids with higher resolutions for the Rayleigh number of $Re=10^6$ in comparison with $Ra=10^5$.

\begin{figure*}
\centering
\includegraphics[width=163 mm]{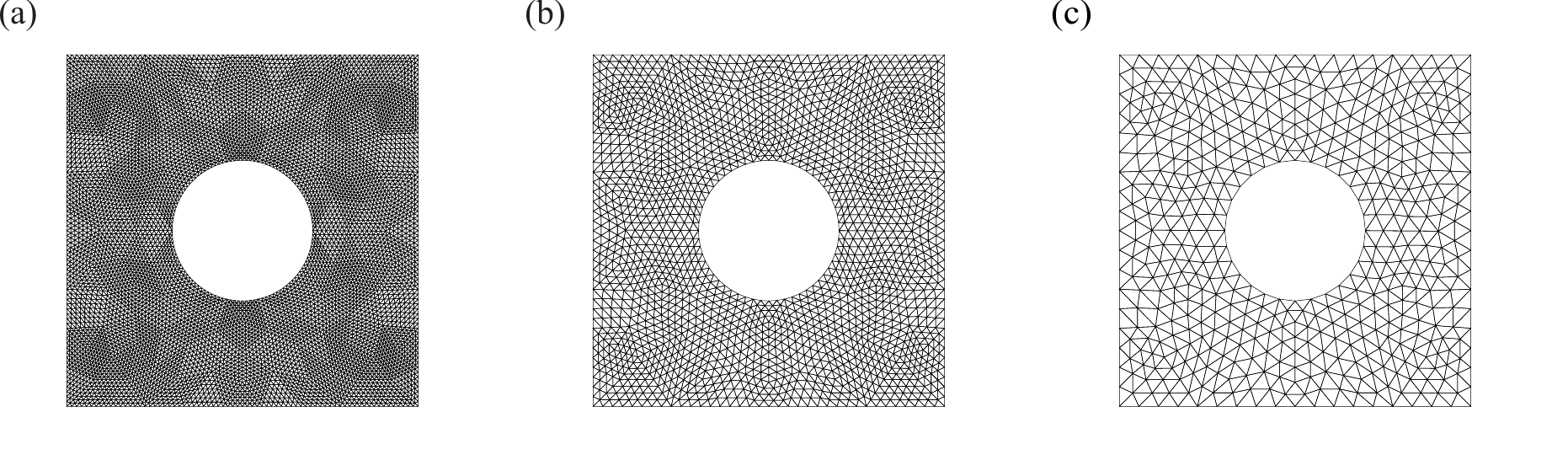}
\caption{The finite element grids utilized for the solution of Poisson's equation in the simulation of natural convection in a square enclosure with a circular cylinder: \textbf{a} Standard resolution ($k=0$), 7560 nodes and 14720 elements; \textbf{b} After one level coarsening ($k=1$), 1940 nodes and 3680 elements; \textbf{c} After two levels coarsening ($k=2$), 510 nodes and 920 elements.}
\label{fig:2}
\end{figure*}

Figures 2--3 depict, respectively, the vorticity lines and isotherms in the buoyancy-driven flow simulation using both the normal and the CGP processes for two different Rayleigh number of $Ra=10^5$ and $Ra=10^6$. The vorticity and thermal fields with one ($k=1$) and two levels ($k=2$) of coarsening agree well with the full fine scale normal computations, and they present significantly more reliable outputs in comparison with the corresponding full coarse scale simulations. This fact, for instance, is noticeable from Fig. 2, when one compares the vorticity lines of the simulations performed on grids with three resolutions of the non-CGP full fine scale ($k=0$ with 14720:14720), CGP ($k=2$ with 14720:920), and non-CGP full coarse scale ($k=0$ with 920:920) at the Rayleigh number of $Ra=10^5$. If one compares the outputs of these three resolutions (mentioned in the last sentence) with each other, but now for the temperature field illustrated in Fig. 3, it can be realized that the efficiency of the CGP technique becomes more pronounced in the vorticity field rather than the temperature field. In the velocity-pressure formulation of the Navier-Stokes equation, the vorticity is a post-processed quantity and is proportional to the spatial gradient of the components of the velocity vector. Hence, CGP is able to maintain excellent accuracy of the velocity gradient as well. We will demonstrate that the fact is true for the temperature gradient when we compute the local Nussult number on the cylinder surface.

\begin{figure*}
\centering
\includegraphics[width=163 mm]{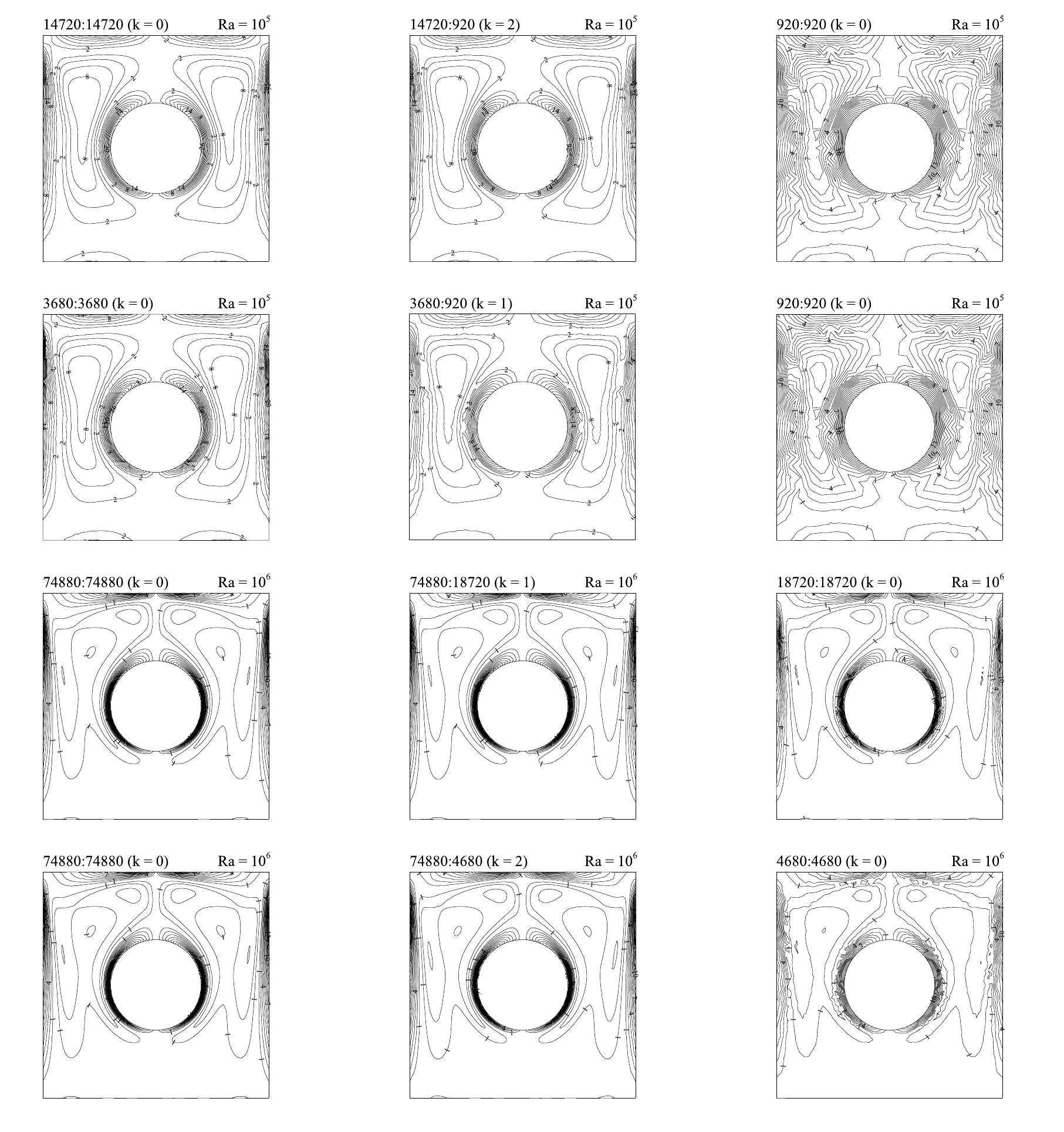}
\caption{Vorticity lines for the buoyancy-driven flow for two different Rayleigh numbers of $Ra=10^5$ and $Ra=10^6$. Labels in the form $M:N$ specify the grid resolution of the advection-diffusion and energy equation solvers, $M$ elements, and the pressure Poisson equation solver, $N$ elements. $k$ indicates the coarsening level.}
\label{fig:2}
\end{figure*}

\begin{figure*}
\centering
\includegraphics[width=163 mm]{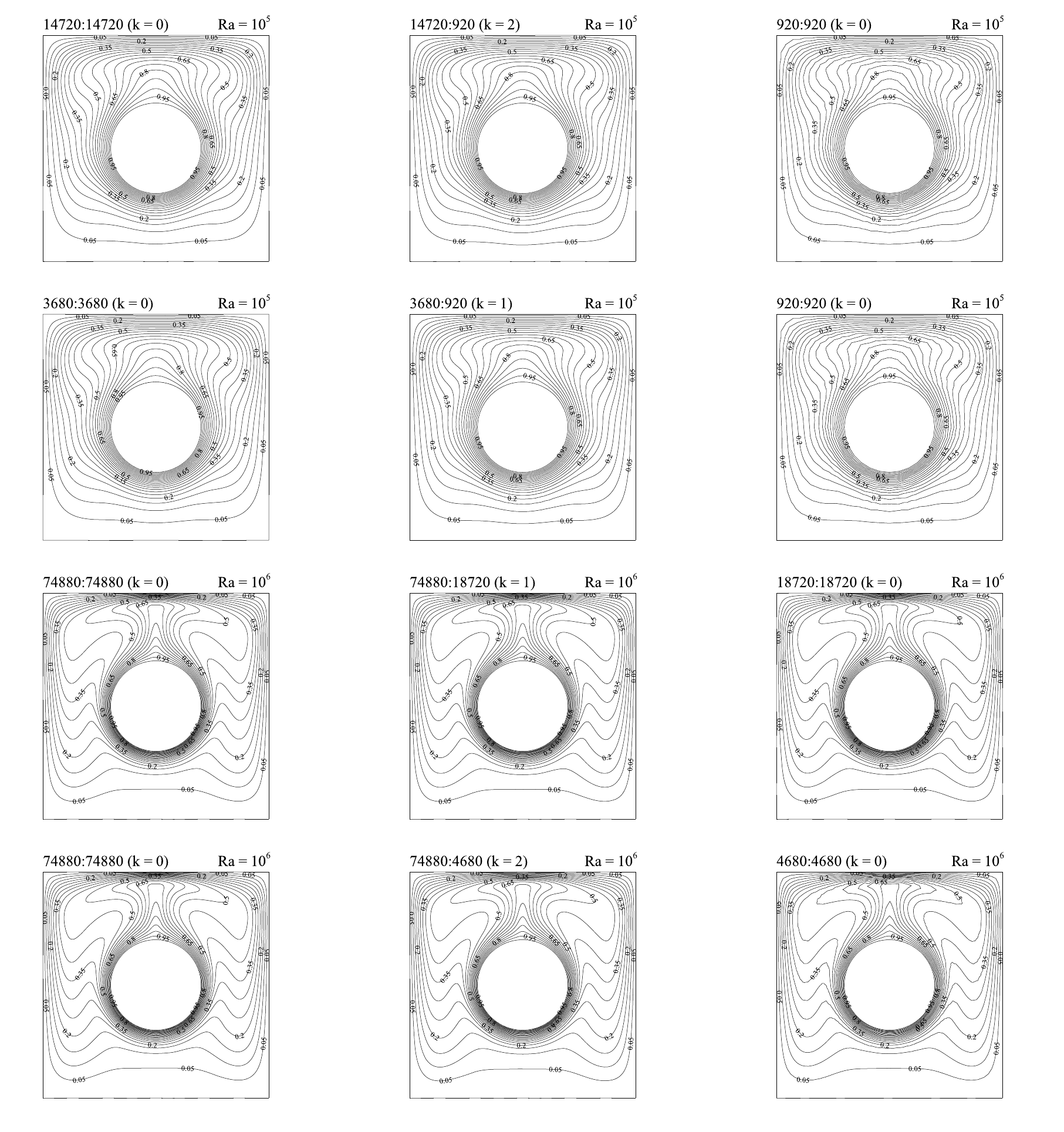}
\caption{Isotherms for the buoyancy-driven flow for two different Rayleigh numbers of $Ra=10^5$ and $Ra=10^6$. Labels in the form $M:N$ indicate the mesh resolution of the advection-diffusion and energy equation solvers, $M$ elements, and the pressure Poisson equation solver, $N$ elements. $k$ shows the coarsening level.}
\label{fig:2}
\end{figure*}

More specifically, Table 1 compares the efficiency and the accuracy of the velocity and temperature fields for the standard approach ($k=0$) and the CGP algorithm ($k=1$, and 2) for the Rayleigh number of $Ra=10^6$. The resulting data captured from the CGP simulations ($k=1$, and 2) is considerably more accurate than the outputs of standard simulations executed on the full coarse scale grid resolutions ($k=0$ with 18720:18720, and $k=0$ with 4680:4680), as can be seen from the computed error norms relative to the simulation performed on the finest mesh ($k=0$ with 74880:74880). Interestingly, the flow field reaches stationary for a relatively equal number of iterations (i.e., time steps) for both the CGP and non-CGP schemes for all the spatial resolutions. However, the CGP method reduces the CPU time per iteration. The maximum achieved speedup is a factor of approximately 2.0. 

\begin{table*}
\centering
\caption{Comparison of relative norm errors for the velocity and temperature fields, total CPU times, number of iterations, and speedup between the standard and CGP algorithms for the buoyancy induced-convection problem at the Rayleigh number of $Ra=10^6$. Grid resolution in the structure of $M:N$ represents the spatial resolution of the advection-diffusion and energy solvers, $M$ elements, and Poisson's equation, $N$ elements.}
\label{tab:5}   
\begin{tabular}{lllllllll}
\hline\noalign{\smallskip}
$k$ & Resolution & $\| \textbf{\textit{u}}\|_{L^\infty(V)}$ & $\| \textbf{\textit{u}}\|_{L^2(V)}$ & $\| \theta \|_{L^\infty(V)}$& $\| \theta \|_{L^2(V)}$ & CPU time (s) & Iterations & Speedup\\
\noalign{\smallskip}\hline\noalign{\smallskip}
0 & 74880:74880 & - & - & - & - & 113922.00 & 3774 & 1.000   \\
1 & 74880:18720 & 1.88739E$-$8 & 1.65035E$-$9 & 4.75281E$-$8 & 4.45308$-$9 & 67023.60 & 3772 & 1.699 \\
2 & 74880:4680 & 1.35787E$-$7 & 9.69478E$-$9 & 4.47647E$-$7 & 2.93357$-$8 & 58342.30 & 3769 & 1.952 \\
\noalign{\smallskip}\hline\noalign{\smallskip}
0 & 18720:18720 & 3.57519E$-$8 & 5.13453E$-$9 & 7.9001E$-$8 & 1.12249$-$8 & 9244.02 & 3772 & 1.000 \\
\noalign{\smallskip}\hline\noalign{\smallskip}
0 & 4680:4680 & 4.32146E$-$7 & 9.70076E$-$8 & 1.0556E$-$6 & 2.21036$-$7 &723.75 & 3775 & 1.000 \\
\noalign{\smallskip}\hline
\end{tabular}
\end{table*}

To more precisely examine the performance of the results produced by CGP, the local Nusselt number ($Nu_{\varphi}$) on the surface of the cylinder for different spatial resolutions for the Rayleigh number of $Ra=10^5$ and $Ra=10^6$ is plotted in Fig. 4. The obtained results by both the CGP and non-CGP methods reveal good agreement with the numerical data reported by Lee et al. [15]. For all levels of coarsening, the CGP approach provides more accurate data compared to that modeled with a full coarse scale computation, with reference to the prediction of full fine scale computations. This trend becomes more obvious for the calculation of the maximum local Nussult number, where it occurs at the bottom surface of the cylinder ($\pi < \varphi < 2\pi$), as can be seen from Fig. 4.

\begin{figure*}
\centering
\includegraphics[width=163 mm]{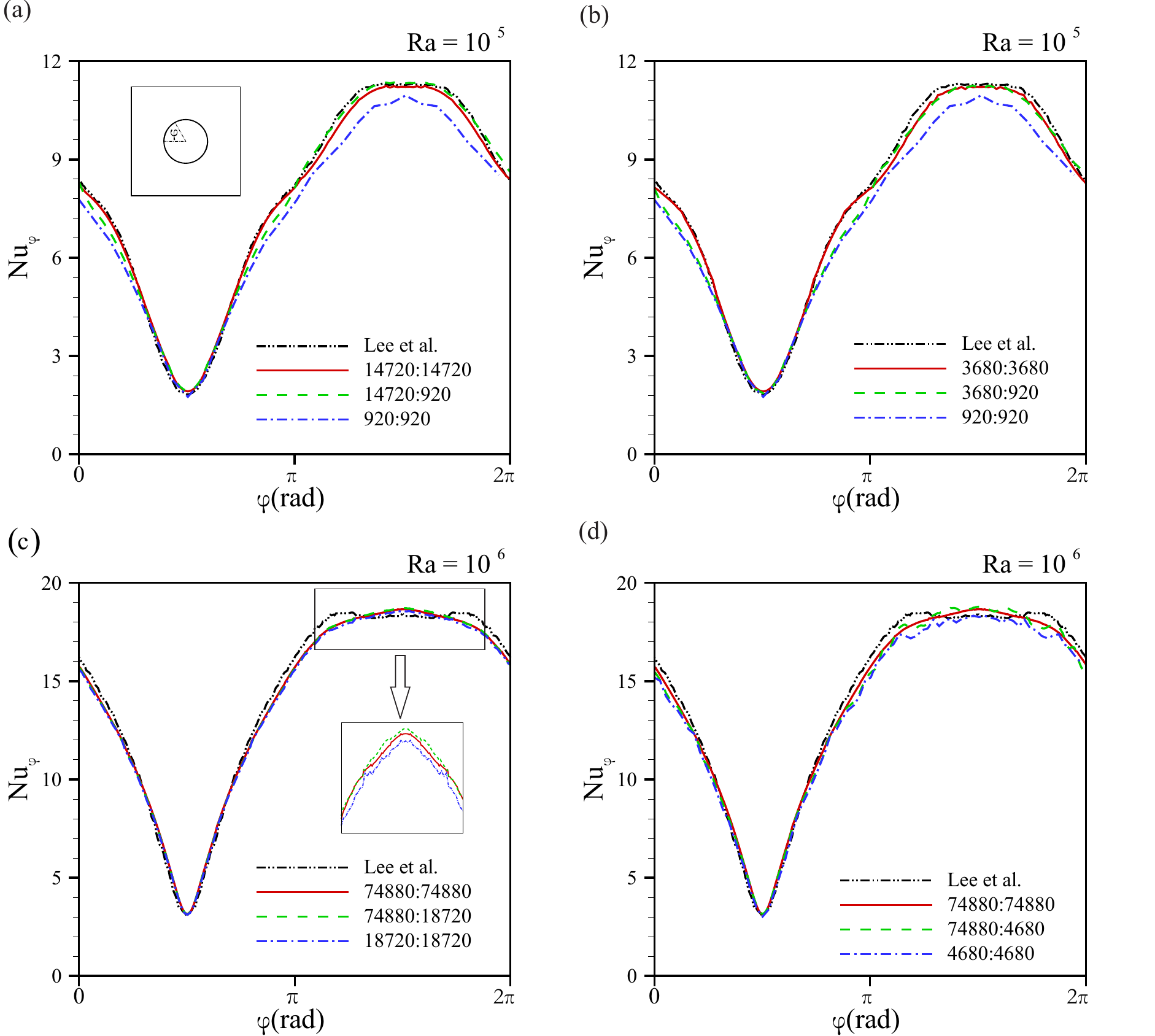}
\caption{Distribution of the local Nusselt number along the surface of the cylinder for the buoyancy-driven 
flow for two different Rayleigh numbers of $Ra=10^5$ and $Ra=10^6$. Resolution in the form of $M:N$ represents the spatial resolution of the advection-diffusion and energy solvers, $M$ elements, and Poisson's equation, $N$ elements.}
\label{fig:2}
\end{figure*}

\subsection{Transport of passive scalars in flows over a circular cylinder}
\label{sec:2}

We consider a rectangular computational domain $V:=[0, 38]\times[0, 32]$. The cylinder is represented by a circle with diameter $d$ in two dimensions. The center of the circle lies at the point (8, 16). At the inflow boundary, we impose a free stream velocity $u_\infty$ perpendicular to the vertical axis, while the outflow boundary is described with a natural Neumann condition
\begin{equation}
\mu \nabla \textbf{\textit{u}} \cdot \textbf{\textit{n}}=0.
\end{equation}

The velocity at the top and bottom of the field is perfectly slipped with the magnitude and direction of $u_\infty$. The circle is considered as a rigid body with no-slip conditions. For the temperature we take boundary conditions from Ref. [17] such that $\theta=\theta_w$ is enforced at the circle, while $\theta=\theta_\infty$ is imposed at the remaining boundaries. The conditions correspond to a problem with constant cylinder temperature. Note that while homogenous natural Neumann conditions are enforced on $\Gamma_N$, temperature Dirichlet conditions are imposed on $\Omega_D$, indicating two different types of boundary conditions at the outflow boundary. The Reynolds number is expressed as
\begin{equation}
Re=\frac{\rho du_\infty }{\mu},
\end{equation}
and the Prandtl number is determined as
\begin{equation}
Pr=\frac{c_p \mu }{\kappa}.
\end{equation}

We determine each point on the circular cylinder surface by the angle $\alpha$ from the negative $x$-axis. Thus, the local Nusselt number on the cylinder surface is formulated as
\begin{equation}
Nu_\alpha=\frac{-d}{\theta_w - \theta_\infty}\frac{\partial \theta}{\partial \textbf{\textit{n}}}\rvert_\alpha.
\end{equation}
The time-averaged Nusselt number per time cycle, $t_p$ , is expressed by
\begin{equation}
Nu=\frac{1}{t_p}\int_{0}^{t_p}Nu_\alpha dt.
\end{equation}
And the time- and space-averaged Nusselt number is calculated by
\begin{equation}
\overline{Nu}=\frac{1}{t_p}\int_{0}^{t_p}(\frac{1}{2\pi}\int_{0}^{2\pi}Nu_\alpha d\alpha) dt.
\end{equation}
The density ($\rho$), free stream velocity ($u_\infty$), specific heat ($c_p$), cylinder temperature ($\theta_w$), and cylinder diameter ($d$) are set to $1.00$; and the temperature at infinity ($\theta_\infty$) is set to 0.00 in the International Unit System. The viscosity ($\mu$) and conductivity ($\kappa$) of the fluid vary to set the Reynolds and Prandtl numbers. A fixed time step of $\delta t=0.05$ s is chosen and we execute the numerical simulations until time $t=150$ s.

The Poisson solver uses the meshes with 108352 nodes and 215680 elements, 27216 nodes and 53920 elements, 6868 nodes and 13480 elements, and 1749 nodes and 3370 elements, respectively, for $k=0$, $k=1$, $k=2$, and $k=3$. Figure 5 shows those grids for $k=2$ and $k=3$.

\begin{figure*}
\centering
\includegraphics[width=168 mm]{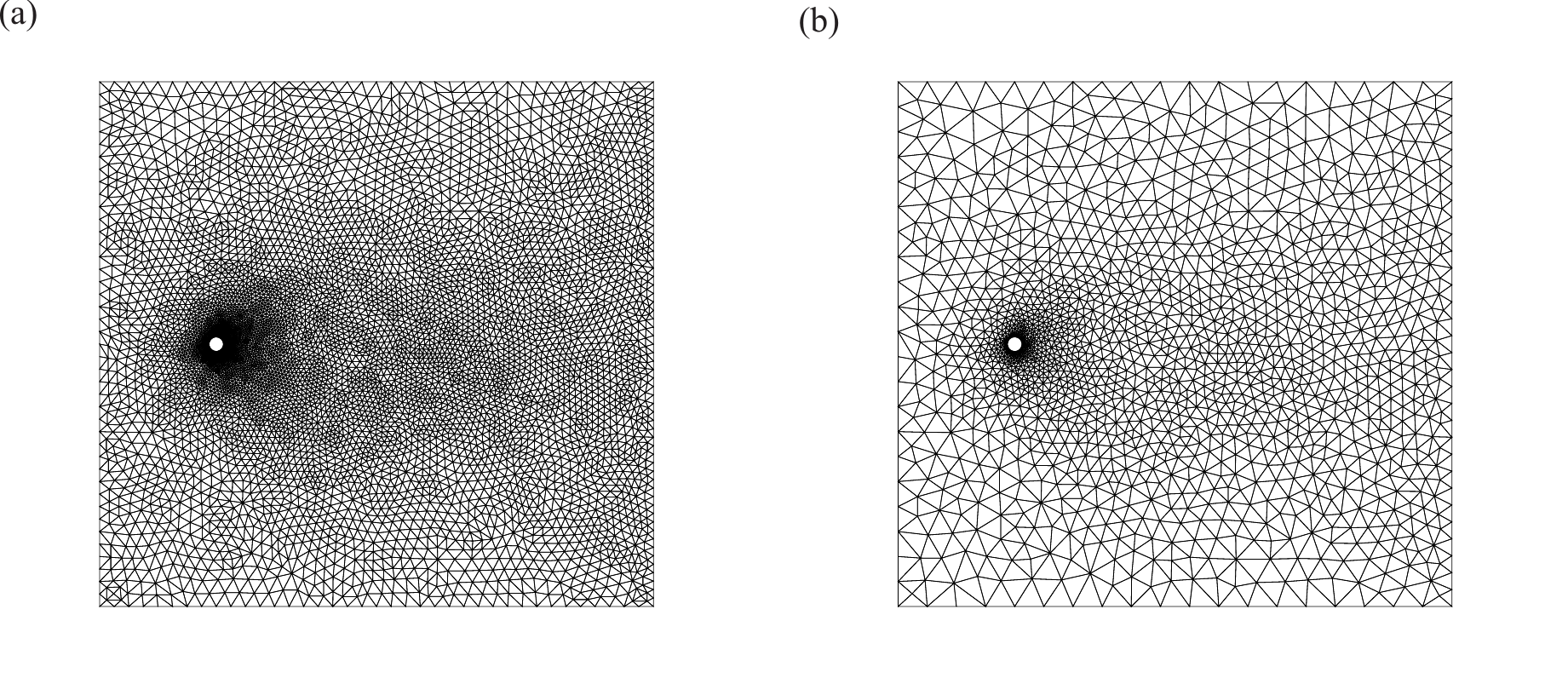}
\caption{Representation of the triangular finite element meshes used for solving Poisson's equation in the simulation of flow over a cylinder. \textbf{a} After two levels coarsening $(k=2)$, 6868 nodes and 13480 elements; \textbf{b} After three levels coarsening $(k=3)$, 1749 nodes and 3370 elements. This figure is reproduced from Ref. [13].}
\label{fig:1}
\end{figure*}

The detailed results related to the velocity field were presented in one of our previous works (see Sect. 3.3 of Ref. [13]) and are not repeated here again.

Table 1 lists the CPU times devoted to each subproblem and speedup factors achieved for the simulations with several spatial resolutions. The most time-consuming component of the simulations with standard resolutions (215680:215680, 53920:53920, 13480:13480, and 3370:3370) is the Poisson equation. Taking the advantages of the CGP method into account, the price of the Poisson equation portion becomes less than 1.2\% only for two levels $(k=2)$ of coarsening. The maximum achieved speedup is a factor of 3.694. In practice, one must solve the linear system of Eq. (22) and Eq. (27) to compute respectively the intermediate velocity field $(\textrm{\~U}_f^{n+1})$ and the temperature field $({\Theta}_f^{n+1})$. From a numerical linear algebra point of view, the $\big(\textbf{M}_v+\delta t \textbf{N}^{n} + \delta t \textbf{L}_v\big)$ and $\big(\textbf{M}_\theta+\delta t \textbf{N}^{n+1} + \delta t \textbf{L}_\theta\big)$ matrices are similar to each other. Hence, the value of the ratio of the computational cost to the number of unknowns is the same for the both systems. The nodal value of the velocity field is twice the nodal value of the temperature field. And this is why for all the simulations with and without the CGP technique, the computational cost of the advection-diffusion equation is roughly twice as much as the cost of the conservation equation of the temperature field. As discussed earlier, the prolongation and restriction operators are constructed based on the idea proposed in Sect. 2.3 of Ref. [13]. Following the data structure introduced in Ref. [13], the numerical expense of the mapping part becomes insignificant, as can be seen from Table 2.

\begin{table*}
\centering
\caption{Total CPU times and their component percentages: the advection-diffusion equation, the pressure Poisson equation, the conservative equation of the passive scalar field, and the mapping function, for the simulation with the Reynolds number of $Re=100$ and the Parndtl number of $Pr=2$. $M:N$ represents the grid resolution of the advection-diffusion and the passive scalar solvers ($M$ elements), and Poisson's equation ($N$ elements).}
\label{tab:1}   
\begin{tabular}{llllllll}
\hline\noalign{\smallskip}
$k$ & Resolution & \%Adv-Dif & \%Poisson & \%Passive & \%Map & CPU (s) & Speedup \\
\noalign{\smallskip}\hline\noalign{\smallskip}
0 & 215680:215680 & 18.25 & 72.61 & 9.14 & 0.000 & 810363.0 & 1.000\\
1 & 215680:53920 & 56.513 & 15.185 & 28.301 & 0.001& 246887.0 & 3.282 \\
2 & 215680:13480 & 65.845 & 1.171 & 32.983 & 0.001 & 221521.0 & 3.658\\
3 & 215680:3370 & 66.524 & 0.102 & 33.373 & 0.001 & 219351.0 & 3.694\\
\noalign{\smallskip}\hline\noalign{\smallskip}
0 & 53920:53920 & 19.976 & 69.962 & 10.062 & 0.000 & 45032.3 & 1.000\\
\noalign{\smallskip}\hline\noalign{\smallskip}
0 & 13480:13480 & 18.248 & 72.486 & 9.266 & 0.000 & 3848.9 & 1.000\\
\noalign{\smallskip}\hline\noalign{\smallskip}
0 & 3370:3370 & 8.889 & 86.922 & 4.189 & 0.000 & 51.1 & 1.000\\
\noalign{\smallskip}\hline
\end{tabular}
\end{table*}

Figure 6 visually compares the temperature fields obtained for $Re=100$ and $Pr=2$ with and without CGP for different grid resolutions at time $t=150$ s. The temperature fields obtained by the CGP procedure for one level $(k=1)$ and two levels $(k=2)$ of the Poisson grid coarsening are close to that simulated with the standard full fine grid resolution (215680: 215680). For three $(k=3)$ levels of coarsening; however, a considerable reduction in the fidelity of the temperature field is observed. Nonetheless, the resulting field of CGP with the 215680:3370 spatial resolution is still better than those that are performed on the standard full coarse grid resolution (3370:3370). Similar observation is reported by Kashefi and Staples [13] for the velocity field (see e.g., Figs. 9--10 of Ref. [13]).

\begin{figure*}
\centering
\includegraphics[width=168 mm]{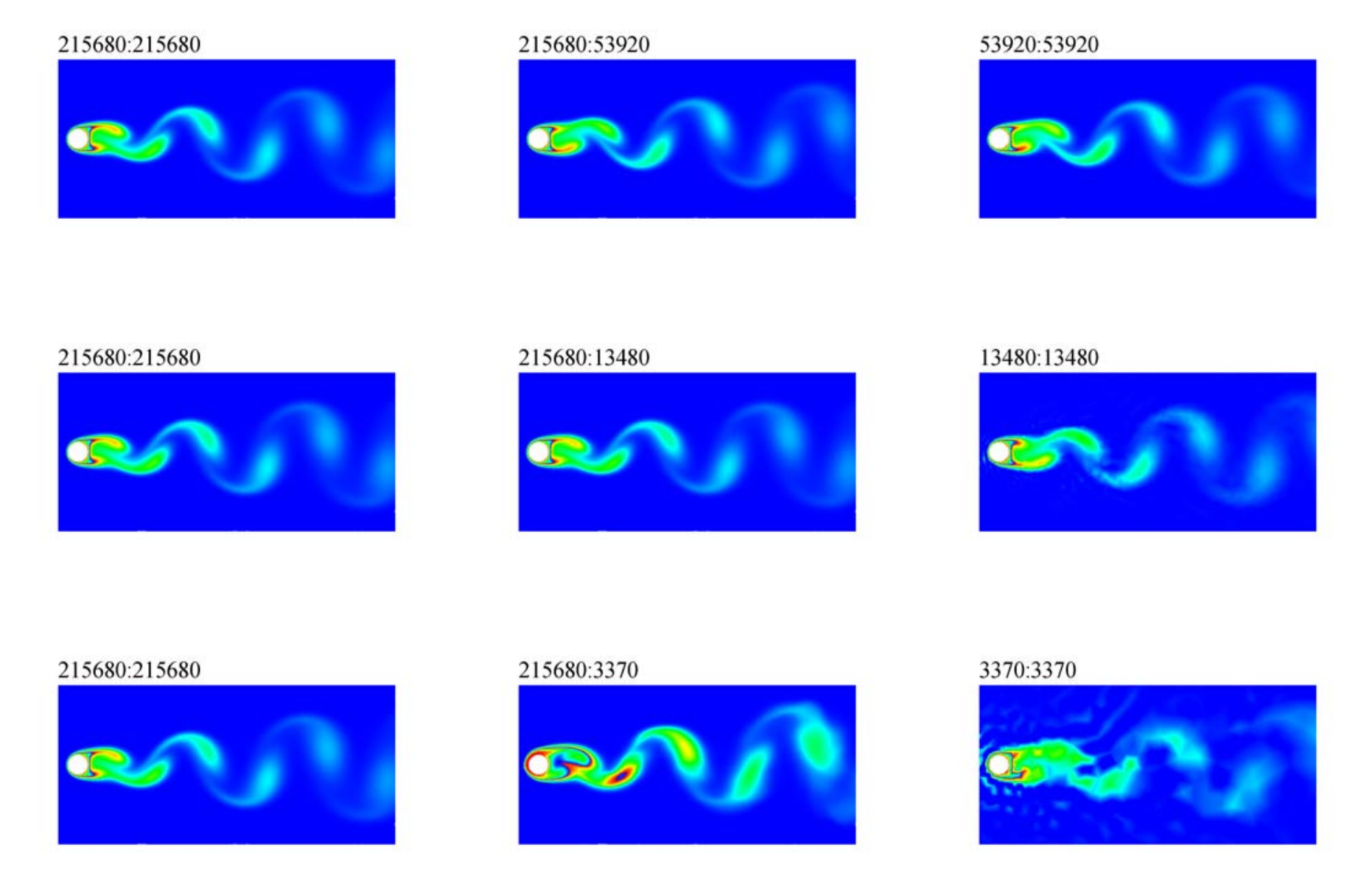}
\caption{Temperature fields for the flow over a circular cylinder for $Re=100$ and $Pr=2$ at $t=150$ s. Labels in the form of
$M:N$ illustrate the grid resolutions of the advection-diffusion and passive scalar fields, $M$ elements, and the pressure field, $N$
elements.}
\label{fig:2}
\end{figure*}

From a general point of view, the spatial discretization of the advection-diffusion domain acts as a lowpass filter on the grid, and the Poisson solver also acts as a pre-filtering process [11]. The CGP procedure specifically uses the belief in order to increase saving in computational time without negatively affecting the properly-resolved velocity field, and consequently the temperature field. A visual demonstration of these effects is displayed in Fig. 7. Figure 7 depicts the temperature distributions along the horizontal centerline in the wake region behind the cylinder for $Re=100$ and $Pr=2$ at time $t=150$ s. While the outputs of the pure coarse grid are contaminated by spurious fluctuations at the end of the fluid domain, these fluctuations are filtered in the temperature field obtained by the CGP framework. 

\begin{figure*}
\centering
\includegraphics[width=165 mm]{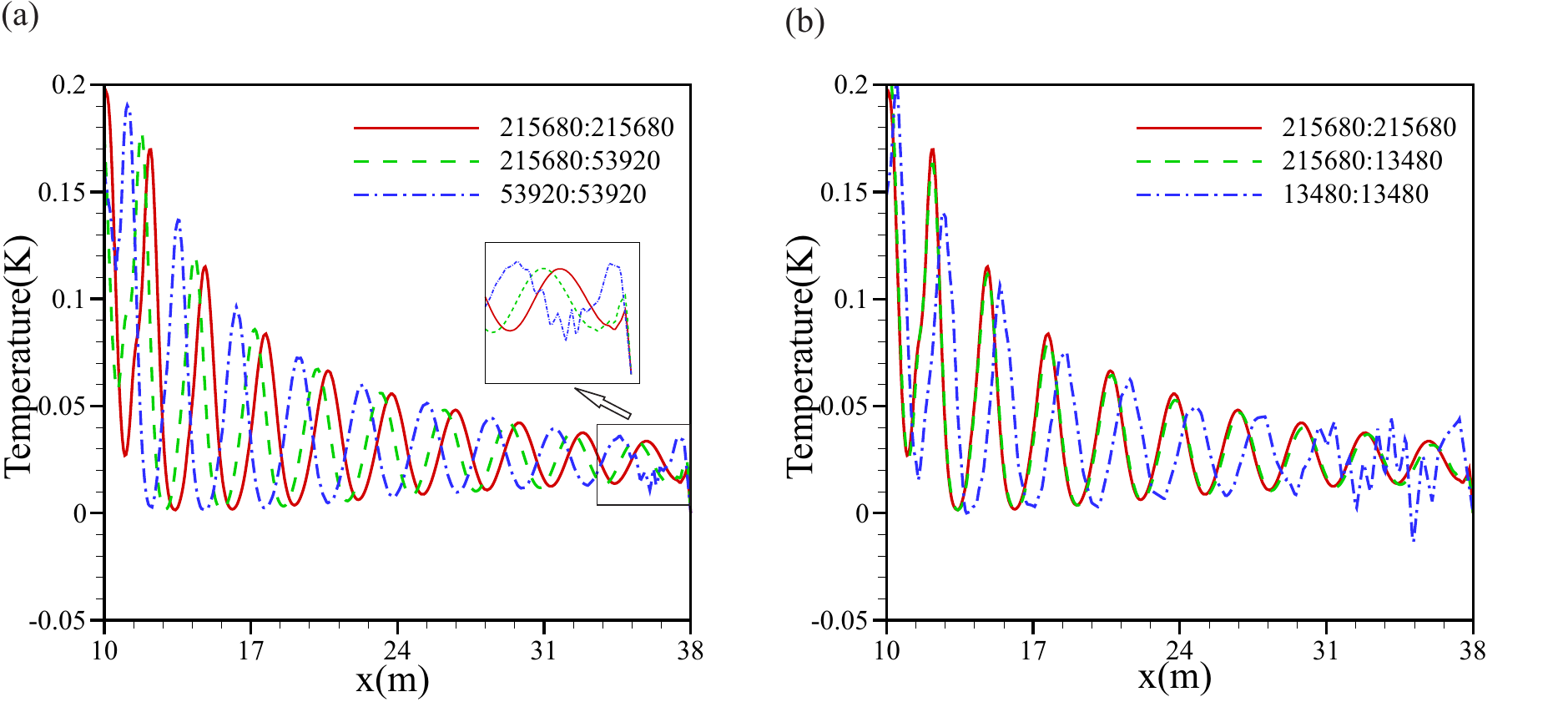}
\caption{ Comparison of the temperature in the wake region of the flow past a cylinder for different grid resolutions for $Re=100$ and $Pr=2$ at $t=150$ s for \textbf{a} $k=0, 1,$ and $0$; \textbf{b} $k=0, 2,$ and $0$. Labels in the form of $M:N$ illustrate the grid resolutions of the advection-diffusion and passive scalar fields, $M$ elements, and the pressure field, $N$ elements.}
\label{fig:12}
\end{figure*}

There could be a phase lag between the velocity outputs of the standard and CGP approaches, depending on the time increment $(\delta t)$ [13]. As can be seen in Fig. 6 and Fig. 7, these phase lags are transmitted from the velocity field into the temperature one.

\begin{figure*}
\centering
\includegraphics[width=165 mm]{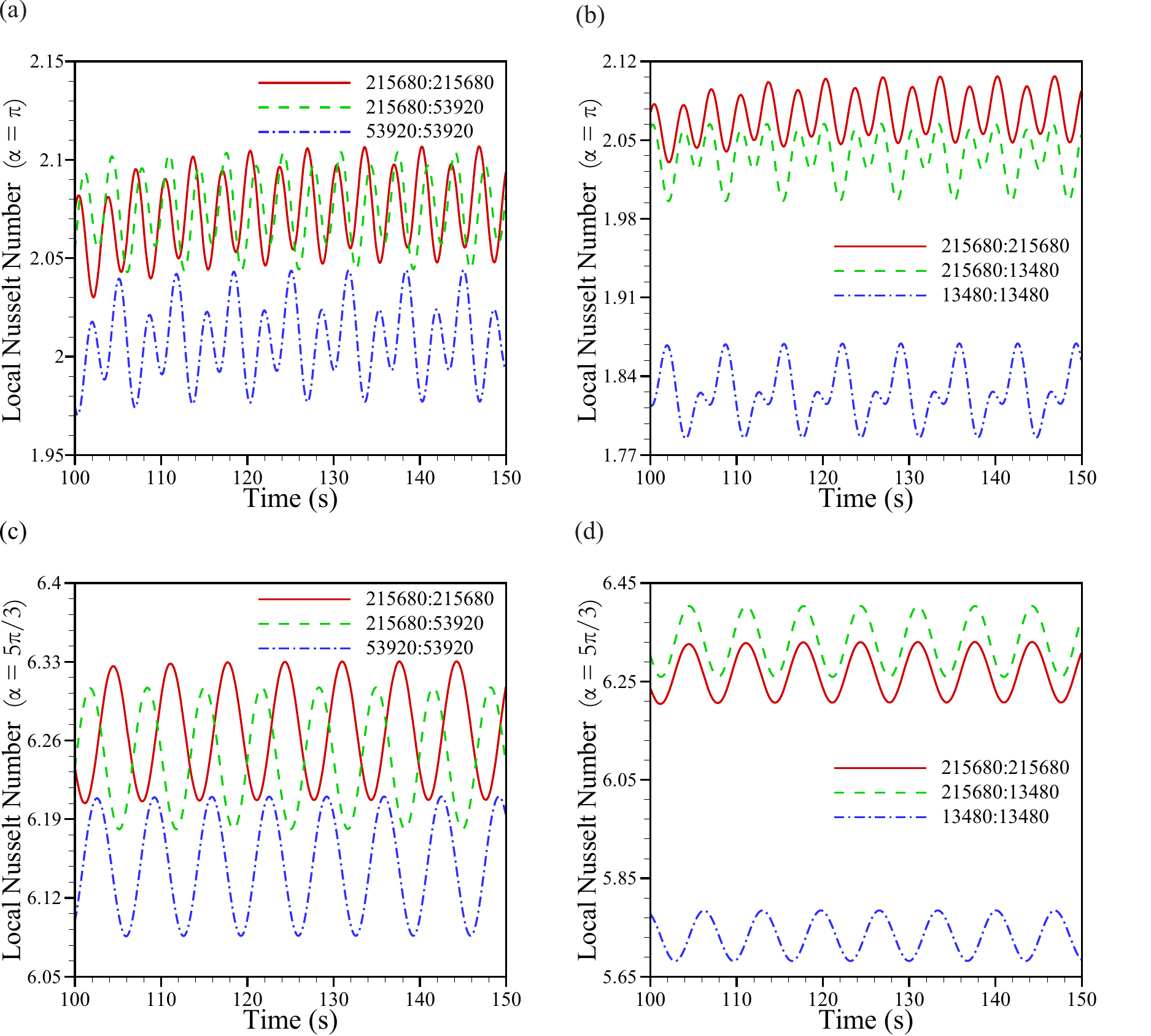}
\caption{The local Nusselt number around the cylinder in the simulation of transport of passive scalars for $Re=100$ and $Pr=0.5$ with and without the CGP algorithm for \textbf{a} one level coarsening $(k=1)$ at angle of $\alpha=\pi$; \textbf{b} two levels coarsening $(k=2)$ at angle of $\alpha=\pi$; \textbf{c} one level coarsening $(k=1)$ at angle of $\alpha=5\pi/3$; \textbf{d} two levels coarsening $(k=2)$ at angle of $\alpha=5\pi/3$. Legends in the form of $M:N$ indicate the spatial resolutions of the advection-diffusion and passive scalar fields, $M$ elements, and the pressure field, $N$ elements.}
\label{fig:13}
\end{figure*}

Figure 8 compares the local Nusselt number $(Nu_\alpha)$ computed for $Re=100$ and $Pr=0.5$ by the standard and CGP algorithms at two different angels of $\alpha=\pi$ and $\alpha=5\pi/3$. For one ($k=1$ with 215680:53920) and two ($k=2$ with 215680:13480) levels of the Poisson grid coarsening, the CGP results are close to the outputs of the full fine scale simulation ($k=0$ with 215680:215680). More importantly, they are significantly more accurate than the local Nusselt number $(Nu_\alpha)$ computed on the full coarse grids ($k=0$ with 53920:53920, and $k=0$ with 13480:13480). For one level $(k=1)$ of mesh coarsening, there is a phase lag between the local Nusselt number $(Nu_\alpha)$ obtained by the standard and CGP simulations at both angels of $\alpha=\pi$ and $\alpha=5\pi/3$ (see Fig. 8a and Fig. 8c). On the other hand, there is no phase lag between the standard and CGP outputs for two levels $(k=2)$ of grid coarsening at these angles (see Fig. 8b and Fig. 8d). Comparing these results with the temperature fields presented in Fig. 7, we experience the same observation. From a mathematical point of view, this phenomena is expected since the local Nusselt number $(Nu_\alpha)$ is proportional to the normal derivative of the temperature variable, and the phase of a continuous bounded oscillatory function gets transmitted to its derivative.

\begin{figure*}
\centering
\includegraphics[width=165 mm]{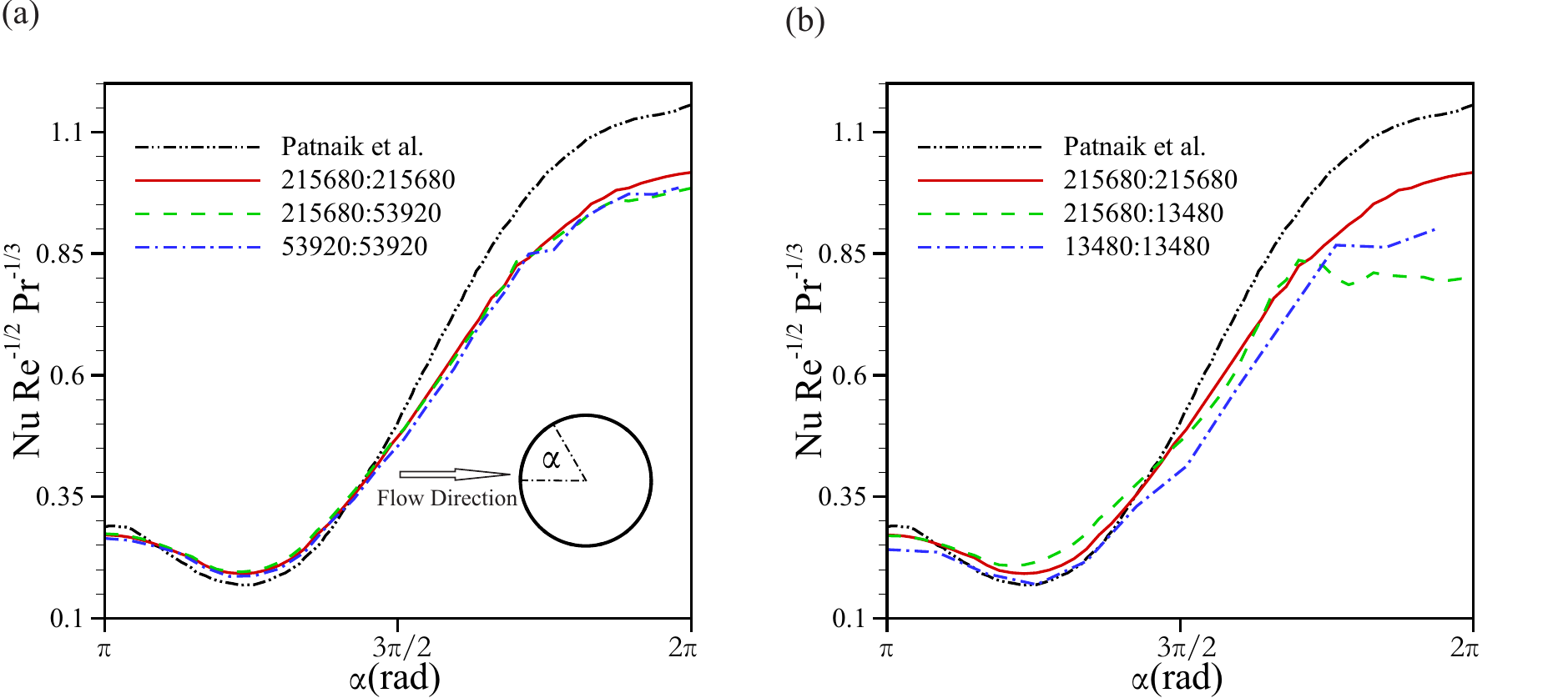}
\caption{ Distribution of time-averaged heat transfer around the cylinder for $Re=100$ and $Pr=0.5$ with and without the CGP algorithm for \textbf{a} one level coarsening $(k=1)$; \textbf{b} two levels coarsening $(k=2)$. Legends in the form of $M:N$ indicate the spatial resolutions of the advection-diffusion and passive scalar fields, $M$   elements, and the pressure field, $N$ elements.}
\label{fig:5}
\end{figure*}

As can be seen in Fig. 8, for two levels $(k=2)$ of the Poisson grid coarsening, the local Nusselt number $(Nu_\alpha)$ predicated by the CGP approach is slightly underestimated at angle of $\alpha=\pi$ in comparison with the outcomes of the full fine scale simulation (215680:215680). We observe, in contrast, an over prediction at angle of $\alpha=5\pi/3$. It is because the architecture of the unstructured coarse grid (with 13480 elements) at these two different zones. Depending on the morphology of a fine grid and the corresponding coarsened grid, the CGP outputs can be slightly over- or under-predicted. We will discuss this issue further in the next paragraph. All in all, the important thing is that although the local Nusselt number $(Nu_\alpha)$ obtained by the CGP computation (215680:13480) is slightly deviated from the full fine scale (215680:215680) result, it is still significantly more accurate than those that are solely computed on the full coarse mesh (13480:13480).

The distribution of the time-averaged Nusselt number $(Nu)$ for $Re=100$  and $Pr=0.5$ on the bottom surface of the cylinder is plotted for the standard and CGP computations with different grid resolutions in Fig. 9. This variable for the simulations with one level ($k=1$ with 215680:53920) and two levels ($k=2$ with 215680:13480) of coarsening agrees well with the data obtained by the standard simulations with the full fine grid resolution ($k=0$ with 215680:215680), approximately from $\alpha=\pi$ to $\alpha=7\pi/4$. For the rest of the bound, even though the local Nusselt number $(Nu_\alpha)$ predicted by the CGP mechanism has a good agreement with the study presented in Ref. [23], a reduction in the accuracy level, in comparison with the full fine scale simulation ($k=0$ with 215680:215680), is observed. This is due to the architecture of the triangular unstructured grids established in this zone. Note that we use the simplest strategy for generating the unstructured grids, while using advanced techniques for mesh generation can significantly affect the performance of GMG tools like CGP.

The data collected in Table 3 demonstrates that the prediction of the time- and space-averaged Nusselt number $(\overline{Nu})$ by the CGP technique is generally reliable and has an excellent agreement with the correlations reported in the literature [24, 25]. By increasing the Prandtl number $(Pr)$, a deviation from the full fine scale computation occurs. In fact at low Prandtl numbers $(Pr)$ the conduction term $(\Delta\theta^{n+1})$ dominates the advection term $((\textbf{\textit{u}}^{n+1}\cdot\nabla)\theta^{n+1})$ in the conservation equation (Eq. (15)) of the temperature. Since a linear mapping function is used here, a higher level of accuracy is obtained at a lower Prandtl number $(Pr)$. Taking the advantages of more advanced data interpolation schemes (see e.g., Ref. [26]) can be a solution to this issue. However, even using the simple extrapolation technique, the outcomes of the CGP configuration are still more accurate than the resulting data captured from the full coarse scale simulations.

\begin{table*}
\centering
\caption{Prediction of the time- and space-averaged Nusslet number $\overline{Nu}$ for the Reynolds number of $Re=100$ for different spatial resolutions. The error percentages are measured with reference to the finest grid resolution. $M:N$ demonstrates the grid resolution of the advection-diffusion and the passive scalar solvers ($M$ elements), and Poisson's equation ($N$ elements).}
\label{tab:2} 
\begin{tabular}{llllll}
\hline\noalign{\smallskip}
$k$ & Resolution & $Pr=0.5$ & \%Error & $Pr=3.0$ & \%Error \\
\noalign{\smallskip}\hline\noalign{\smallskip}
0 & 215680:215680 & 4.4162 & -- & 9.5254 & -- \\
1 & 215680:53920 & 4.3723 & 0.994 & 9.1850 & 3.574 \\
2 & 215680:13480 & 4.2350 & 4.103 & 7.5469 & 20.770 \\
\noalign{\smallskip}\hline\noalign{\smallskip}
0 & 53920:53920 & 4.3102 & 2.400 & 8.7447 & 8.195 \\
\noalign{\smallskip}\hline\noalign{\smallskip}
0 & 13480:13480 & 4.0242 & 8.876 & 7.3741 & 22.585 \\
\noalign{\smallskip}\hline\noalign{\smallskip}
& Fand [24] & 4.8070 & -- & 8.2285 & -- \\
\noalign{\smallskip}\hline\noalign{\smallskip}
& Churchill and Bernstein [25] & 3.5908 & -- & 8.7382 & -- \\
\noalign{\smallskip}\hline
\end{tabular}
\end{table*}

\section{Conclusions and future directions}
\label{sec:1}
In this article, we used for the first time the CGP multigrid scheme to reduce the computational cost for obtaining a numerical solution to the temperature field. In order to examine the performance of CGP, two standard test cases were investigated: Natural convection in a square enclosure with a circular cylinder, and transport of passive scalars in flows past a circular cylinder with constant temperature. The speedup factors ranged approximately from 1.7 to 3.7. The minimum speedup occurred in the thermally-driven flow problems with velocity Dirichlet boundary conditions. However, the maximum speedup belonged to the flow past a cylinder with stress free boundary conditions. A similar conclusion was reported by Kashefi and Staples [13]. For one and two levels of the Poisson grid coarsening, the isotherms and vorticity lines for the buoyancy-driven flow, the structure of the von Karman street for the passive scalars, and generally the heat transfer coefficients were in excellent agreement with those simulated using pure fine grid computations. However, only a reasonable level of accuracy was obtained for three levels of the Poisson mesh coarsening.

The objective of our future research is to perform a comparison between the CGP approach with one level of coarsening ($k=1$) and the standard finite element algorithm with Taylor-Hood mixed finite elements \textbf{P}$_2$/\textbf{P}$_1$ (see e.g., Ref. [19]). From a grid resolution point of view, for an assumed number of grid points of the velocity component, the Poisson solver utilizes a space with an equal pressure node numbers, discretized using either the CGP ($k=1$) method or Taylor-Hood elements. In this sense, a detailed investigation of the similarity/difference between these two concepts may introduce novel mapping functions for the CGP tool.
 
%and subsection a unique label (see Sect.~\bibitem{Ref31}).
%\paragraph{Paragraph headings} Use paragraph headings as needed.

\begin{acknowledgements}
AK would like to thank Dr. Peter Minev and Dr. Omer San for helpful discussions.
\end{acknowledgements}


\begin{thebibliography}{}

\bibitem{Ref1}
Chorin, A. J. Numerical solution of the Navier-Stokes equations. \textit{Mathematics of computation}, \textbf{22}, 745--762 (1968) 
\bibitem{Ref2}
Temam, R. Sur l'approximation de la solution des équations de Navier-Stokes par la méthode des pas fractionnaires (II). \textit{Archive for Rational Mechanics and Analysis}, \textbf{33}, 377--385 (1969)
\bibitem{Ref3}
Timmermans, L., Minev, P. and Van De Vosse, F. An approximate projection scheme for incompressible flow using spectral elements. \textit{International journal for numerical methods in fluids}, \textbf{22}, 673--688 (1996)
\bibitem{Ref4}
Guermond, J., Minev, P. and Shen, J. An overview of projection methods for incompressible flows. \textit{Computer methods in applied mechanics and engineering}, \textbf{195}, 6011--6045 (2006)
\bibitem{Ref5}
Reusken, A. Fourier analysis of a robust multigrid method for convection-diffusion equations. \textit{Numerische Mathematik}, \textbf{71}, 365--397 (1995)
\bibitem{Ref6}
Filelis-Papadopoulos, C. K., Gravvanis, G. A. and Lipitakis, E. A. On the numerical modeling of convection-diffusion problems by finite element multigrid preconditioning methods. \textit{Advances in Engineering Software}, \textbf{68}, 56--69 (2014)
\bibitem{Ref7}
Gupta, M. M., Kouatchou, J. and Zhang, J. A compact multigrid solver for convection-diffusion equations. \textit{Journal of Computational Physics}, \textbf{132}, 123--129 (1997) 
\bibitem{Ref8}
Gupta, M. M., Kouatchou, J. and Zhang, J. Comparison of second-and fourth-order discretizations for multigrid Poisson solvers. \textit{Journal of Computational Physics},  \textbf{132}, 226--232 (1997)
\bibitem{Ref9}
Zhang, J. Fast and high accuracy multigrid solution of the three dimensional Poisson equation. \textit{Journal of Computational Physics}, \textbf{143}, 449--461 (1998)
\bibitem{Ref10}
Lentine, M., Zheng W. and Fedkiw, R. A novel algorithm for incompressible flow using only a coarse grid projection. \textit{In: ACM Transactions on Graphics}, 114--122 (2010)
\bibitem{Ref11}
San, O. and Staples, A. E. A coarse-grid projection method for accelerating incompressible flow computations. \textit{Journal of Computational Physics}, \textbf{233}, 480--508 (2013)
\bibitem{Ref12}
Jin, M., Liu, W. and Chen, Q. Accelerating fast fluid dynamics with a coarse-grid projection scheme. \textit{HVAC\&R Research}, \textbf{20}, 932--943 (2014) 
\bibitem{Ref13}
Kashefi, A. and Staples, A. E. A finite-element coarse-grid projection method for incompressible flow simulations. \textit{Advances in Computational Mathematics}, \textbf{44}, 1063--1090 (2018)
\bibitem{Ref14}
Kashefi, A. Coarse grid projection methodology: A partial mesh refinement tool for incompressible flow simulations. \textit{Bulletin of the Iranian Mathematical Society} , in press (2019)
\bibitem{Ref15}
Lee, J. M., Ha, M. Y. and Yoon, H. S. Natural convection in a square enclosure with a circular cylinder at different horizontal and diagonal locations. \textit{International Journal of Heat and Mass Transfer}, \textbf{53}, 5905--5919 (2010)
\bibitem{Ref16}
Kim, J. and Moin, P. Transport of passive scalars in a turbulent channel flow. \textit{Proc. 6th Symp. on Turbulent Shear Flows}, France, 521--526 (1987) 
\bibitem{Ref17}
Karniadakis, G. Numerical simulation of forced convection heat transfer from a cylinder in crossflow. \textit{International Journal of Heat and Mass Transfer}, \textbf{31}, 107--118 (1988)
\bibitem{Ref18}
Wanner, G. and Hairer, E. \textit{Solving ordinary differential equations II}, Springer-Verlag, Berlin (1991)
\bibitem{Ref19}
Reddy, J. N. \textit{An introduction to the finite element method}, McGraw-Hill, New York (1993)
\bibitem{Ref20}
Saad, Y. and Schultz, M. H. GMRES: A generalized minimal residual algorithm for solving nonsymmetric linear systems. \textit{SIAM Journal on scientific and statistical computing}, \textbf{7}, 856--869 (1986)
\bibitem{Ref21}
Van der Vorst, H. A. \textit{Iterative Krylov methods for large linear systems}, Cambridge University Press, Cambridge (2003)
\bibitem{Ref22}
Geuzaine, C. and Remacle, J. F. Gmsh: A 3‐D finite element mesh generator with built‐in pre‐and post‐processing facilities. \textit{International Journal for Numerical Methods in Engineering}, \textbf{79}, 1309--1331 (2009) 
\bibitem{Ref23}
Patnaik. V., Narayana, A. and Seetharamu, K. Numerical simulation of vortex shedding past a circular cylinder under the influence of buoyancy. \textit{International Journal of Heat and Mass Transfer}, \textbf{42}, 3495--3507 (1999)
\bibitem{Ref24}
Fand, R. M. Heat transfer by forced convection from a cylinder to water in crossflow. \textit{International Journal of Heat and Mass Transfer}, \textbf{8}, 995--1010 (1965)
\bibitem{Ref25}
Churchill, S. W. and Bernstein, M. A correlating equation for forced convection from gases and liquids to a circular cylinder in crossflow. \textit{Trans. ASME, Ser. C, J. Heat Transfer}, \textbf{99}, 300--306 (1977)
\bibitem{Ref26}
Amsallem, D. and Farhat, C. Interpolation method for adapting reduced-order models and application to aeroelasticity. \textit{AIAA Journal}, \textbf{46}, 1803--1813 (2008)

% etc
\end{thebibliography}


\end{document}